**Title:** Ultra-broadband bright light emission from a one-dimensional inorganic van der Waals material

**Author Names:** Fateme Mahdikhany[1*], Sean Driskill[1*], Jeremy G. Philbrick[1], Davoud Adinehloo[2], Michael R. Koehler[3], David G. Mandrus[4-6], Takashi Taniguchi[7], Kenji Watanabe[8], Brian J. LeRoy[1], Oliver L.A. Monti[1,9], Vasili Perebeinos[2], Tai Kong[1,3], John R. Schaibley[1]

**Author Addresses:**

[1]Department of Physics, University of Arizona, Tucson, Arizona 85721, USA

[2]Department of Electrical Engineering, State University of New York at Buffalo, Buffalo, New York, 14260

[3]IAMM Diffraction Facility, Institute for Advanced Materials and Manufacturing, University of Tennessee, Knoxville, TN 37920

[4]Department of Materials Science and Engineering, University of Tennessee, Knoxville, Tennessee 37996, USA

[5]Materials Science and Technology Division, Oak Ridge National Laboratory, Oak Ridge, Tennessee 37831, USA

[6]Department of Physics and Astronomy, University of Tennessee, Knoxville, Tennessee 37996, USA

[7]Research Center for Materials Nanoarchitectonics, National Institute for Materials Science,  1-1 Namiki, Tsukuba 305-0044, Japan

[8]Research Center for Electronic and Optical Materials, National Institute for Materials Science, 1-1 Namiki, Tsukuba 305-0044, Japan

[9]Department of Chemistry and Biochemistry, University of Arizona, Tucson, Arizona 85721, USA

[*]Fateme Mahdikhany and Sean Driskill are equal contributors to this work and designated as co-first authors.
**Corresponding Author:** John Schaibley, johnschaibley@arizona.edu

**Abstract**

One-dimensional (1D) van der Waals materials have emerged as an intriguing playground to explore novel electronic and optical effects. We report on inorganic one-dimensional $SbPS_4$ nanotubes bundles obtained via mechanical exfoliation from bulk crystals. The ability to mechanically exfoliate $SbPS_4$ nanobundles offers the possibility of applying modern 2D material fabrication techniques to create mixed-dimensional van der Waals heterostructures. We find that $SbPS_4$ can readily be exfoliated to yield long (> 10 µm) nanobundles with thicknesses that range from of 1.3 - 200 nm. We investigated the optical response of semiconducting SbPS$_4$ nanobundles and discovered that upon excitation with blue light, they emit bright and ultra-broadband red light with a quantum yield similar to that of hBN-encapsulated MoSe$_2$. We discovered that the ultra-broadband red light emission is a result of a large ~1 eV exciton binding energy and a ~200 meV exciton self-trapping energy, unprecedented in previous material studies. Due to the bright and ultra-broadband light emission, we believe that this class of inorganic 1D van der Waals semiconductors has numerous potential applications including on-chip tunable nanolasers, and applications that require ultra-violet to visible light conversion such as lighting and sensing. Overall, our findings open avenues for harnessing the unique characteristics of these nanomaterials, advancing both fundamental research and practical optoelectronic applications.

**Main Text:**

**Introduction**

One-dimensional (1D) van der Waals (vdW) materials exhibit a 1D vdW gap, meaning that their crystal structure is composed of nanotubes with terminated chemical bonds. Carbon nanotubes and transition metal dichalcogenide nanotubes have been the subject of intense research for decades and are known to exhibit exceptional electronic and optical properties, including strongly bound excitons (electron-hole pairs)[1-5]. There has also been significant interest in highly anisotropic materials such as transition metal trichalcogenides that exhibit quasi-1D crystal structures that are known to exhibit superconductivity and charge density wave phases[6]. However, these well-known 1D vdW materials are typically obtained either through chemical synthesis or chemical vapor deposition and have limited compatibility with modern 2D material fabrication and transfer techniques.

SbPS$_4$ is a direct bandgap semiconductor first reported by D'ordyai et al.[7]. Its crystal structure, shown in Figure 1a, indicates that it is a true 1D van der Waals material. Hollow tubes with a diameter of ~1.3 nm, formed by trivalent antimony atoms in connection to [PS$_4$]$^{3+}$ tetrahedra, are terminated by sulfur atoms on their exterior, and weakly bonded to each other via the vdW interaction. Previously, direct synthesis[8] and vapor transport[9] method were used to obtain pure crystalline SbPS$_4$. The vapor transport method allows crystal synthesis directly on fluorine doped tin oxide glass or silicon substrates.

**Results and Discussion**

In this paper, we use the direct synthesis method to first obtain bulk single crystalline SbPS4. Specifically, bulk crystals are obtained by melting stoichiometric amounts of the starting elements in a vacuum sealed silica tube at 650°C and then slowly cooling down to room temperature. Scotch tape mechanical exfoliation method, which is widely used in the 2D community, was then used to exfoliate bulk crystals down to nanoscale for measurements. This sample preparation approach gives a 100% sample yield and is compatible with well-established 2D nanodevice fabrication techniques[10].

Remarkably, we were able to isolate long (> 10 μm) nanobundles of $SbPS_4$ with thicknesses on the order of 1.3 - 200 nm. Figure 1b shows an optical microscope image of a typical nanobundle on a 285 nm $SiO_2$/Si substrate. The dashed box shows an area of the sample where the sample topography was directly imaged with atomic force microscopy (AFM) (Figure 1b, top left inset). The top right inset of Figure 1b shows a topography line scan measured at the location indicated by the white arrow in the AFM image, where the $SbPS_4$ has a height of 80 nm and width of ~700 nm.

$SbPS_4$ nanobundles were optically investigated using photoluminescence (PL) and Raman spectroscopy. Specifically, we used a 0.6 NA objective to focus laser light to a ~1 μm spot on the sample, and then collected the light emitted/scattered from the nanobundle with the same objective in the reflection geometry. The sample was placed in an optical cryostat allowing for measurements down to 4 K. Figure 1c shows a PL spectrum from an isolated nanobundle when excited with 3.06 eV (405 nm) at a sample temperature of 4 K. The PL spectrum consists of an asymmetric peak, shown in Figure 1c. The spectrum was fit with a sum of two Voigt peak functions, where the low energy and high energy peaks are centered around 1.36 eV and 1.56 eV, respectively. The low energy peak is consistently brighter than the high energy peak across samples with varying thickness (see Supplementary Figure 1). We assign the higher energy peak to free excitons with a 1 eV binding energy, while the lower energy peak is attributed to self-trapped excitons with a ~200 meV trapping energy corresponding to very strong electron-phonon interactions (see discussion below). Figure 1d depicts the proposed energy level diagram for SbPS$_4$, where a high energy laser excites free electrons and holes which then rapidly and non-radiatively relax to form free and self-trapped excitons that emit PL.

Consistent with previous reports on semiconducting carbon nanotubes, the SbPS$_4$ PL exhibits a linear polarization dependence[11]. To probe this, we used a linear polarizer and an achromatic half-wave plate to vary the excitation and detection polarization axis simultaneously, measuring the PL that is co-linearly polarized with the excitation laser. Figure 2a is an optical microscope image of a nanobundle and defines the detection axis as well as the bundle axis. The polarization dependent PL intensity is shown in Figure 2b. PL intensity is maximized when the detection axis is either parallel or anti-parallel to the bundle axis. When the detection axis and bundle axis are perpendicular, the PL intensity is ~68% of the maximum. We were not able to calculate the ratio of the parallel to perpendicularly polarized excitonic PL directly. Instead, we theoretically calculated both the parallel and perpendicular absorption coefficient spectra (see Supplementary Figure 2) which provides a measure of the expected excitonic polarization dependence. We then integrated these two spectra and found that the integrated perpendicular absorption coefficient is 60% of the parallel, in excellent agreement with our measurement of 68%. Supplementary Note 1 and Supplementary Figure 2 provide a detailed theoretical analysis of the SbPS$_4$ polarization dependence.

To understand the physical origins of the SbPS$_4$ emission, we measured the PL as a function of excitation energy, temperature and polarization angle. PL spectra as a function of excitation energy were measured using a broadly tunable supercontinuum laser source (Figure 2c). We observe a clear onset of the PL intensity when the excitation photon energy is above 2.75 eV, which is consistent with previous reports of the 2.57 eV bandgap of the SbPS$_4$[8]. The PL intensity rapidly increases above 2.75 eV excitation energy, reaching a maximum near 3 eV. We note that the two-peak structure of the PL spectrum (highlighted in Figure 1c) persists across the different excitation energies. The PL intensity also has a strong temperature dependence. PL is detectible at room temperature (see Supplementary Figure 3), but increases significantly at temperatures below 60 K. PL spectra for sample temperatures ranging from 4 K to 100 K are shown in Figure 2d (with an excitation energy of 3.06 eV). The PL intensity exhibits a maximum around 30 K, and then decreases to half the maximum by ~70 K. The relative intensities of the two-peak structure are largely unaffected by temperature up to 60 K, after which the PL becomes more difficult to measure with high signal-to-noise ratio.

The non-monotonic temperature dependence is attributed to the existence of a dark state that traps excitons at low temperature. Here, we assume that photoexcited states reach thermal equilibrium on the radiative lifetime time scale. Therefore, the radiative decay rate as a function of temperature can be calculated using[12] :

$$I_{PL}(T) \propto \frac{1}{Z(T)} \sum_\mu \exp(-E_{\mu 0}/k_B T),$$  (1)

where index $\mu$ runs over all bright exciton states with momentum $q = 0$, $E_{\mu 0}$ are the energies of two excitons , and $Z(T)$ is the partition function of all possible excited states with finite momentum in all bands $\mu$: $Z(T) = \sum_{\mu q} \exp(-E_{\mu q}/k_B T)$. In the case of a single emitting exciton band, the problem reduces to evaluating the partition function. For a 1D parabolic single exciton band $Z(T) \propto T^{1/2}$, leading to the celebrated result[13] that $I_{PL}(T) \propto T^{-1/2}$. However, a realistic excitonic bandstructure involves multiple bands, thereby giving rise to a complicated form of the partition function[14]. To account for this, we fit our experimentally measured temperature dependence to the following empirical expression motivated by Ref.[12]:

$$\frac{I_{PL}(T)}{I_{PL}(T_m)} = \left(\frac{T_m}{T}\right)^\alpha \exp\left(\frac{T-T_m}{T}\right),$$  (2)

where $T_m$ and $\alpha$ are two fitting parameters. Parameter $T_m$ depends on the bright-dark exciton energy splitting[12]. We find $T_m = 17$ K and $\alpha = 1.4$, which suggests that bright-dark exciton splitting is about 2 meV, consistent with that in carbon nanotubes of similar diamters[15], which reinforces our interpretation of the PL spectra from the large binding energy excitons.

The exciton binding energy, $E_b$, in 1D structures follows an exciton scaling law[16]:

$$E_b = A_b R^{\alpha-2} m^{\alpha-1} \varepsilon^{-\alpha},$$  (3)

where $m$ is the exciton reduced mass (in units of free electron mass $m_e$), $R$ is the radius of the tube (in nm), and the empirical parameters $A_b = 24.1$ eV and $\alpha = 1.4$ were determined previously[16]. Using $R = 0.58$ nm and $m = 0.99\, m_e$ from our DFT calculation of the effective masses of the hole ($m_{hole} = 2.43\, m_e$) and electron ($m_{electron} = 1.67\, m_e$) in SbPS$_4$ we can rationalize experimentally observable binding energy of 1 eV using a single parameter fit $\varepsilon = 4.5$, which is a typical value for inorganic materials, such as SbPS$_4$.

We propose that the lower energy PL peak originates from strong exciton-phonon coupling, an effect that is related to the soft SbPS$_4$ lattice. To understand this strong exciton-phonon coupling, we explored the optically active vibrational modes of the nanobundles by Raman spectroscopy and calculated theoretically electron-phonon couplings by density functional theory as implemented in the Quantum Espresso software[17]. In the Raman experiments, the nanobundles were excited using 20 μW with a 2.3 eV photon energy laser. The experimentally measured Raman spectrum is the yellow curve shown in Figure 3a. Our Raman spectra are consistent with previous reports[9,18]. Theoretical calculations of Gamma-point phonons weighted by the strength of electron-phonon couplings to the top of the valence and bottom of the conduction band edges (blue curve of Figure 3a) show eleven peaks representing the phonon modes between Raman shifts of 100 cm$^{-1}$ and 600 cm$^{-1}$, and the experimental measurements (yellow curve) match eight of the eleven calculated phonon modes, providing strong agreement between theory and experiment. The peak observed at 520 cm$^{-1}$ corresponds to the Raman peak of Si.

To justify the strongly self-trapped exciton model, we first performed a systematic theoretical analysis of the electronic structure of SbPS$_4$ and followed this by a calculation of the polaron binding energy for electrons and holes. The details of the electronic structure calculation are given in Supplementary Note 2, and Supplementary Figure 4. The resulting total density of states (DOS) and projected DOS for the Sb, P, and S atoms are shown in Figure 3b.

To estimate polaron binding energy in SbPS$_4$, we calculated contributions to the polaron binding from individual phonons in a single unit cell of 76 atoms. We used DFT to calculate the electron-phonon coupling ($C_\nu$) to the band edge states and the effective atomic spring constant ($k_\nu$) for each of the $\nu = 1 \dots 225$ vibration modes at the Gamma point. A simple model suggests that each vibrational mode contributes to the distortion amplitude ($C_\nu/k_\nu$) and to the polaron binding energy:

$$E_P = \sum_\nu \frac{C_\nu^2}{k_\nu} \qquad (4)$$

We utilized the Phonopy package to obtain the phonon energies and corresponding eigenvectors[19]. The response of eigenvalues to normal phonon distortions was used to calculate $C_\nu$ and the total energy change to calculate the spring constant $k_\nu$. The summation of each phonon contribution to the binding energy suggests that the polaron binding energies for electrons and holes in SbPS$_4$ are 51.7 meV and 82.2 meV, respectively. The phonon modes weighted by the polaron binding energy contributions according to Eq. (4) are shown in Figure 3a. The agreement with the Raman signal intensities, which are also proportional to the strength of the electron-phonon coupling, is remarkable. The total calculated binding energy for the electron and hole (51.7 meV + 82.2 meV = 133.9 meV) is in remarkable agreement (within less than a factor of two) with the experimental measurement of 200 meV, the energy difference between the two PL peaks. The contribution of phonons to polaron energies in Figure 3a indicates that there is no particular phonon which contributes primarily to the exciton binding, but rather many phonons over the entire range contribute non-negligibly, justifying our more rigorous approach to investigating self-trapping.

Finally, in order to probe the exciton dynamics, we carried out time resolved PL measurements using time correlated single photon counting. The nanobundles were excited with short < 300 ps optical pulses with an energy of 3.06 eV (405 nm). Figure 4a shows time resolved PL measured at three specific PL detection energies. For the purpose of determining the lifetime, we used a triexponential fit on each curve displayed as solid lines in Figure 4a. We identify three distinct time scales: $T_1$ ~10-40 $\pm$ 15 ns, $T_2$ ~ 700 $\pm$ 110 ns, and $T_3$ ~7000 $\pm$ 400 ns. We attribute $T_1$ and $T_2$ to the free and self-trapped exciton lifetime respectively. The extremely long microsecond timescale ($T_3$) is currently attributed to a dark state (mentioned above), and will be the subject of future investigations.

**Conclusion**

In summary, we have shown that long nanobundles of $SbPS_4$ can be mechanically exfoliated, and exhibit bright PL with a high quantum yield. The PL exhibits an exceptionally broad spectrum spanning nearly 1.2-1.9 eV which arises from strongly bound self-trapped excitons with quantum yield comparable to encapsulated $MoSe_2$ (Supplementary Figure 5). We note that a previous report on SbPS4 grown via vapor transport reported a PL peak centered around 1.2-1.9 eV, which was attributed to defect-related emission, and a 2.5 eV PL peak which is absent in our samples (Supplementary Figure 3). However, in our work, we carried out a systematic experimental and theoretical investigation that shows that the 1.5 eV PL peak originates from strongly bound self-trapped excitons. This model is fully supported by our theoretical analysis that predicts a 1 eV exciton binding energy as well as a ~100-200 meV self-trapping energy in remarkable agreement with our experimental finding. Additionally, consistent measurements across various samples, without any notable shift in the energy center, further demonstrate that the PL is intrinsic rather than extrinsic. We emphasize that the ultra broadband and efficient nature of the light emission makes these nanoscale materials attractive for future optoelectronic applications including broadband tunable lasers, and on chip optical interconnects. The absorption of UV light and efficient emission in the visible spectrum also makes these materials attractive as "phosphors" in UV to visible light conversion applications such as sensitive high energy particle detection and high efficiency fluorescent lighting. Notably, these materials can be excited at 405 nm and emit in the red spectrum. This is particularly significant given the long-standing challenge of developing red phosphors that can be efficiently pumped in the blue range, with 450 nm being the ideal for solid-state lighting. We envision that this material can be exfoliated down to the single 1D chain, opening up novel opportunities to explore intrinsic 1D electronic physics. Due to their van der Waals nature, we believe that this class of materials is ideal to realize novel mixed 1D-2D material heterostructures using modern polymer transfer techniques well established in the 2D material community.

In the final stages of preparation of this manuscript, we became aware of another work reporting on the optical response of $SbPS_4$ grown via vapor transport[9].

**Methods**

**$SbPS_4$ Crystal Growth and Characterization**

Starting elements, antimony shots (Alfa Aesar, 99.999%), phosphorus pieces (Beantown Chemical, 99.999%) and sulfur pieces (Thermo Scientific, 99.999%) were mixed in a stoichiometric ratio, and vacuum sealed in a silica tube. The sealed tube was then heated up to $650°$ C and slowly cooled to room temperature. Powder x-ray diffraction of the synthesized $SbPS_4$ was measured using a Bruker D8 diffractometer (See Supplementary Figure 6a). The nanotube structure was also confirmed with high resolution transmission electron miscopy (Supplementary Figure 6b).

**Mechanical Exfoliation**

Mechanical exfoliation was used to separate $SbPS_4$ bundles from their bulk material. Bulk $SbPS_4$ was placed onto the adhesive side of scotch tape. The tape was then folded onto itself until the $SbPS_4$ was distributed across its surface. A second piece of tape was used to pick up and deposit bundles onto oxygen plasma treated SiO$_2$/Si substrates. The tape was peeled off the substrate at an approximate rate of 2 mm per minute.

**Optical Measurements**

SbPS$_4$ bundles were measured in an optical cryostat in vacuum. Laser light was focused into the cryostat using a 40X/0.6 NA microscope objective. The spectra were detected using a grating spectrometer and a cooled CCD camera (Andor). Due to CCD efficiency limitations, PL intensities below 1.25 eV could not be reliably detected. We used the CCD efficiency curve to correct the PL spectra for the spectrometer response. To avoid burning the bundles, laser power did not exceed 5 μW for ~400 nm wavelength excitation.

For low-temperature PL and polarization-resolved PL, the sample temperature was fixed at 4 K. Bundles were excited with a continuous wave 405 nm diode laser which has been further spectrally filtered using a 450 nm short-pass filter.

For the photoluminescence excitation (PLE) and temperature dependent PL measurements, the sample was excited with a tunable, broadband supercontinuum laser (NKT SuperK FIANIUM) source which was passed through a tunable optical filter and a 650 nm short-pass filter. The spectra were filtered using a 700 nm long-pass filter before entering the spectrometer. For PLE, the sample temperature was 4 K while the excitation wavelength varied from 405 nm to 650 nm. For temperature dependence, the excitation wavelength was fixed at 405 nm while the sample temperature varied from 4 K to 100 K.

To measure the polarization-dependent PL, we employed a fixed polarizer along with an achromatic half-wave plate situated between the beam splitter and the sample. This setup enabled us to control the excitation and detection polarization axes simultaneously.

For Raman spectroscopy, the sample temperature was 4 K. A 532 nm diode laser with 20 μW power was used to excite the sample. The Raman signal was separated from the back reflection by using a steep edge 532 nm long-pass filter before being detected. Phonon modes below a Raman shift of 100 cm$^{-1}$ could not be observed due to the limitations of the filter.

In the time-resolved PL measurements, we used a broadband, pulsed laser (NKT SuperK FIANIUM) with a tunable optical filter, a time-correlated single photon counting system (PicoHarp 300), and a single photon avalanche detector. The laser repetition rate was set to 150 kHz at 405 nm wavelength. A 700 nm long-pass filter rejected the back reflected laser. The sample temperature was 4 K. A spectrometer was used to isolate specific detection energies.

**Supplementary Material:**

The supplementary material for this article contains PL spectra from three additional samples, the theoretical analysis of the polarization dependent oscillator strength, the theoretical analysis of the electronic structure, a comparison of the $SbPS_4$ PL with monolayer $MoSe_2$, example x-ray diffraction data, and a transmission electron microscope image.

**Acknowledgments:**

This work is mainly supported by the AFOSR Grant Nos. FA9550-22-1-0312, FA9550-22-1-0220, and FA9550-21-1-0219. P.V acknowledge support from AFOSR Grant No. FA9550-22-1-0312. D.G.M. acknowledges support from the Gordon and Betty Moore Foundation's EPiQS Initiative, Grant GBMF9069. K.W. and T.T. acknowledge support from the JSPS KAKENHI (Grant Numbers 21H05233 and 23H02052) and World Premier International Research Center Initiative (WPI), MEXT, Japan. B.J.L. and J.R.S. acknowledges additional support from NSF Grant Nos. DMR-2003583, ECCS-2054572, and the Army Research Office under Grant no. W911NF-20-1-0215. J.R.S. acknowledges additional support from AFOSR Grant No. FA9550-20-1-0217. B.J.L. acknowledges support from NSF Grant No. ECCS- 2122462. T.K. acknowledges support from the University of Arizona.

**Data Availability:**

The data that support the findings of this study are available from the corresponding author upon reasonable request.

**Code Availability:**

Upon request, authors will make available any previously unreported computer code or algorithm used to generate results that are reported in the paper and central to its main claims.

**Contributions:**

J.R.S. and T.K. conceived and supervised the project. F.M. and S.D. fabricated the structures and performed the experiments, assisted by J.G.P.. F.M. and S.D. and D.A. analyzed the data with input from J.R.S., V.P., T.K., O.L.A.M. and B.J.L. T.K. and J.G.P. grew and characterized the SbPS$_4$. M.R.K. and D.G.M provided MoSe$_2$ crystals. T.T. and K.W. provided hBN crystals. V.P. and D.A. provided theoretical support in interpreting the results. F.M., S.D., J.R.S., D.A., V.P. wrote the paper with input from T.K., B.J.L. and O.L.A.M. All authors discussed the results.

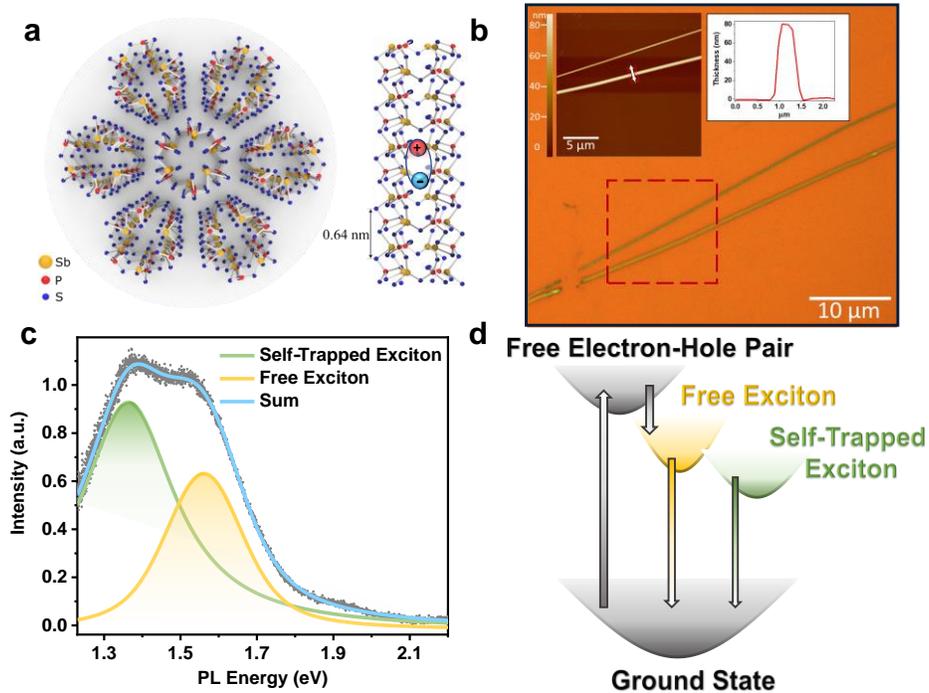

**Figure 1: Structure and photoluminescence of mechanically exfoliated SbPS₄ nanotube bundles.** (a) Depiction of the SbPS₄ lattice structure parallel and perpendicular to the tube axis. Triangular unit cells are held together at the Sb and S bonds. (b) Optical image and AFM topography (left inset) of SbPS₄ bundles. The right inset offers a line cut of the topography, the location of which is displayed with an arrow in the AFM topography, where the SbPS₄ has a height of 80 nm and width of ~700 nm. (c) PL spectrum of an SbPS₄ bundle when excited with a 405 nm (3.06 eV) laser at 4 K. The asymmetric peak can be deconstructed to two separate peaks. The low-energy, bright peak, highlighted in green, corresponds to self-trapped exciton emission, and the high-energy peak associated with free exciton emission shown in yellow. (d) Energy level diagram of the SbPS₄ PL. Nonradiative relaxations are shown with gray arrows. A free electron and hole are excited from the ground state. The electron-hole pair is then bound to form a free exciton. Strong exciton-lattice coupling distorts the lattice, resulting in a lower energy self-trapped exciton. Radiative transitions occur during the relaxation of the free exciton or the self-trapped exciton to the ground state.

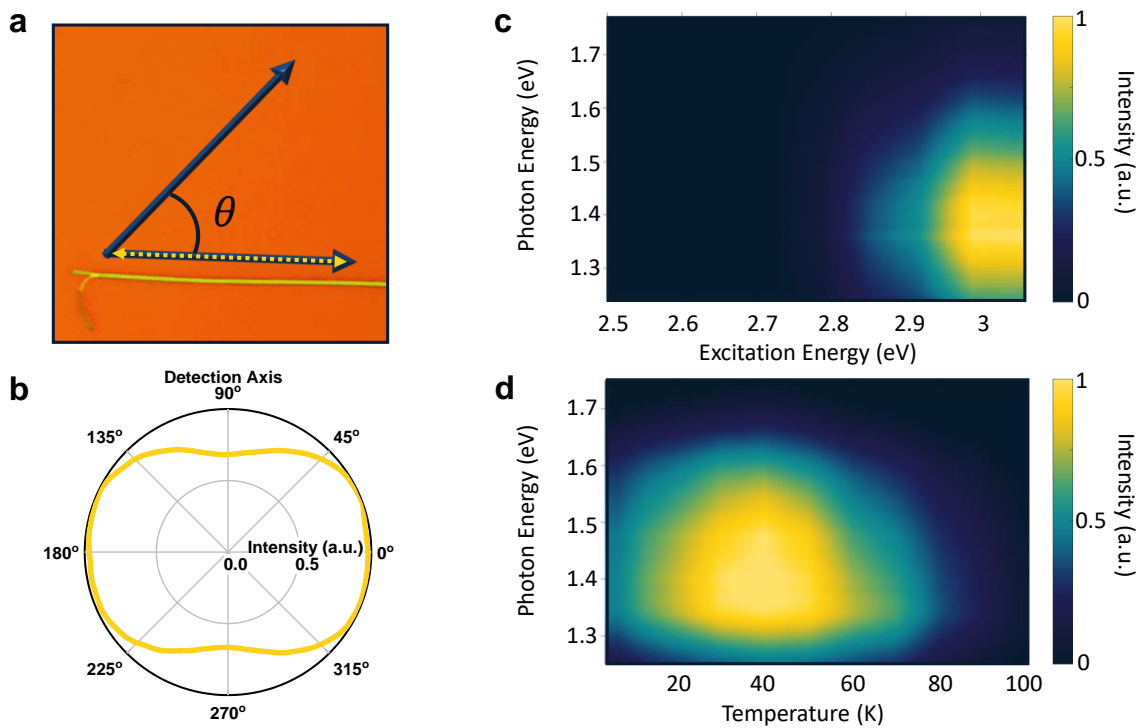

**Figure 2: PL dependence on excitation polarization, energy and sample temperature.** (a) Depiction of the detection and bundle axes used in polarization-dependent PL measurements overlaid on an optical image of the nanobundle. (b) Linear polarization dependence of SbPS$_4$ PL spectra. (c) PL as a function of excitation photon energy. (d) PL intensity as function of sample temperature reveals the intensity of PL is maximum near 40 K.

**a**

**b**

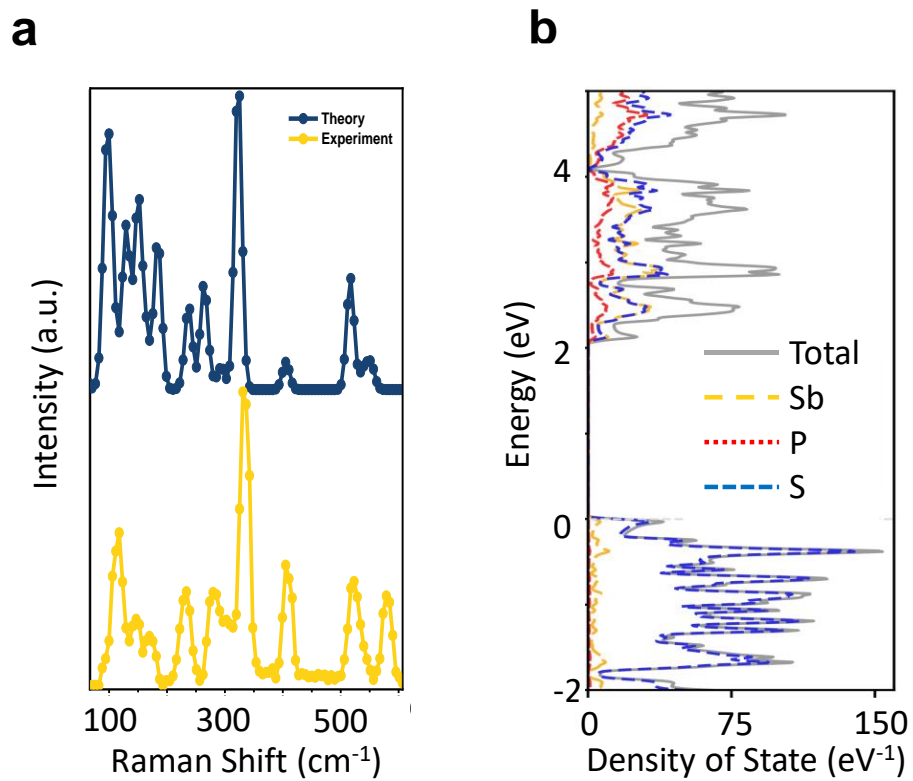

**Figure 3: 1D vibrational and electronic structure.** (a) DFT vibrational modes weighted by the strength of electron-phonon coupling (blue) and experimental Raman spectra (yellow) between 100 cm⁻¹ and 600 cm⁻¹ Raman shift. (b) The computed total density of states, detailing the contributions from Sb (yellow), P (red), and S (blue) atoms.

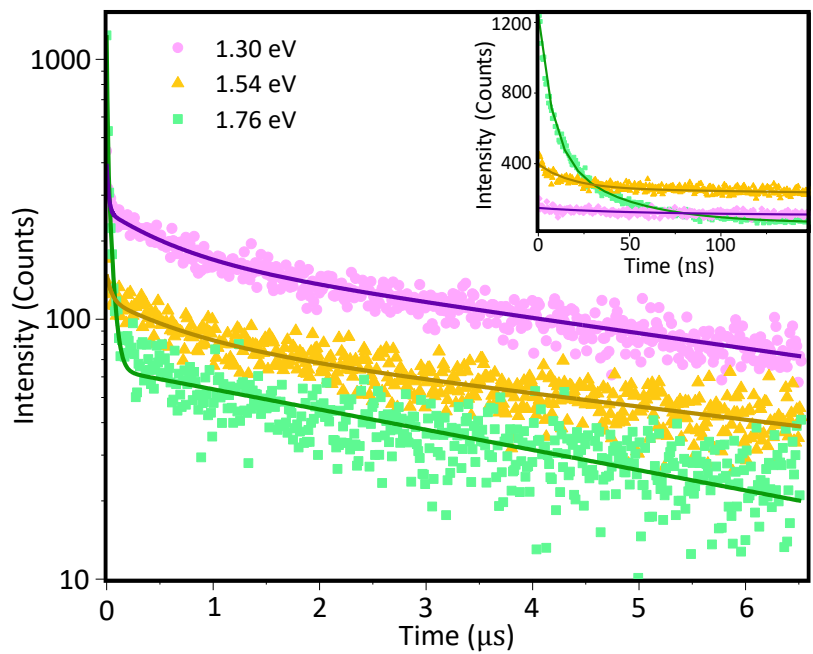

**Fig 4: 1D exciton dynamics.** Illustration of the time-resolved PL for three distinct photon energies, represented on a semi-logarithmic scale. Each curve is fit to a triexponential decay function, revealing three unique radiative decay lifetimes. The inset shows the first 150 ns.

# Supplementary Information for

**Title:** Ultra-broadband bright light emission from a one-dimensional inorganic van der Waals material

**Author Names:** Fateme Mahdikhany[1*], Sean Driskill[1*], Jeremy G. Philbrick[1], Davoud Adinehloo[2], Michael R. Koehler[3], David G. Mandrus[4-6], Takashi Taniguchi[7], Kenji Watanabe[8], Brian J. LeRoy[1], Oliver L.A. Monti[1,9], Vasili Perebeinos[2], Tai Kong[1,3], John R. Schaibley[1]

**Author Addresses:**

[1]Department of Physics, University of Arizona, Tucson, Arizona 85721, USA

[2]Department of Electrical Engineering, State University of New York at Buffalo, Buffalo, New York, 14260

[3]IAMM Diffraction Facility, Institute for Advanced Materials and Manufacturing, University of Tennessee, Knoxville, TN 37920

[4]Department of Materials Science and Engineering, University of Tennessee, Knoxville, Tennessee 37996, USA

[5]Materials Science and Technology Division, Oak Ridge National Laboratory, Oak Ridge, Tennessee 37831, USA

[6]Department of Physics and Astronomy, University of Tennessee, Knoxville, Tennessee 37996, USA

[7]International Center for Materials Nanoarchitectonics, National Institute for Materials Science, 1-1 Namiki, Tsukuba 305-0044, Japan

[8]Research Center for Functional Materials, National Institute for Materials Science, 1-1 Namiki, Tsukuba 305-0044, Japan

[9]Department of Chemistry and Biochemistry, University of Arizona, Tucson, Arizona 85721, USA

[*]Fateme Mahdikhany and Sean Driskill are equal contributors to this work and designated as co-first authors.

**Corresponding Author:** John Schaibley, johnschaibley@arizona.edu

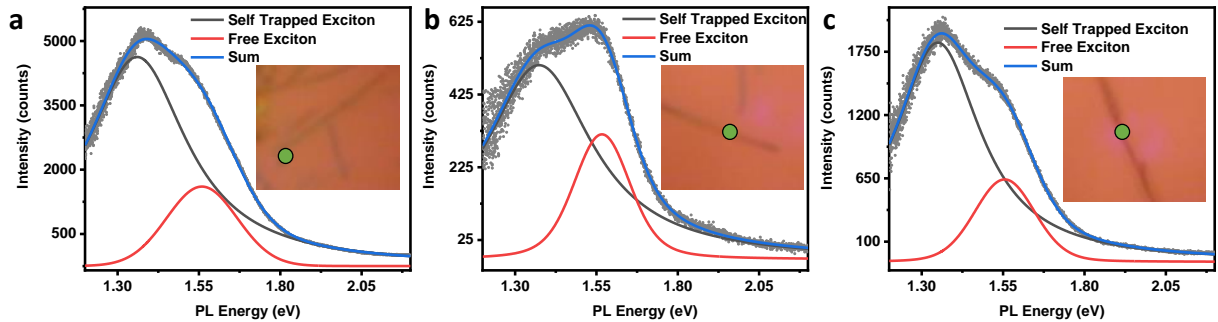

**Supplementary Figure 1: 2 Peak PL Fitting From Multiple Nanobundles.** (a)-(c) Example spectra of SbPS$_4$ for three different nanobundles, fitting the data with two Voigt functions, represented in black and red. The inset optical image displays the excitation spot as viewed through a 40x / 0.6 NA microscope objective. As it is shown, even with variations in the intensity ratio between free and self-trapped excitons, the emission from the self-trapped exciton consistently is brighter than that of the free exciton. Furthermore, the center of energy for both the self-trapped and free excitons remains unchanged across all samples.

**Supplementary Note 1: Theory of anisotropic SbPS₄ absorption.**

The assessment of optical absorption properties plays a vital role in determining the suitability of materials for optoelectronic applications. The optical absorption coefficient, which quantifies the material's ability to absorb incident light, can be determined by calculating the dielectric function[1],

$$\alpha(\omega) = \frac{\omega \varepsilon_2}{nc} = \sqrt{2}\frac{\omega}{c}\left(\sqrt{\varepsilon_1^2 + \varepsilon_2^2} - \varepsilon_1\right)^{1/2},$$

where $\omega$ represents the frequency of the incident light, $c$ is the speed of light, $n$ stands for the refractive index, while $\varepsilon_1$ and $\varepsilon_2$ correspond to the real and imaginary components of the dielectric function, respectively. Parallel to the tube and perpendicular to the tube optical absorption coefficient of SbPS₄ are shown in Supplementary Figure 2. It shows that the intensity of the perpendicular optical absorption coefficient is 60% of the parallel (along the nanowire) optical absorption coefficient, which is consistent with the experimental data and indicates the anisotropic behavior of the SbPS₄.

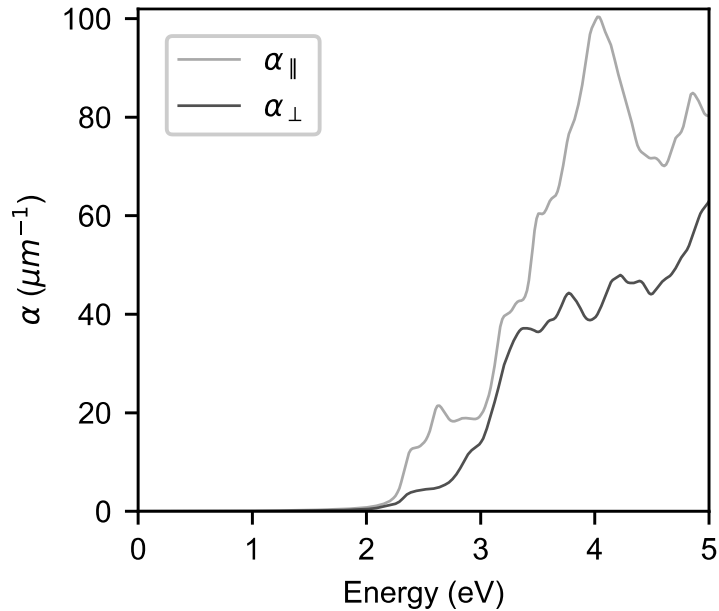

**Supplementary Figure 2: Optical absorption coefficient of SbPS₄.** Absorption coefficient of SbPS₄ parallel to the tube ($\alpha_\parallel$) and perpendicular ($\alpha_\perp$) to the tube. Integral of the perpendicular optical absorption coefficient is 60% of the parallel (along the nanowire) optical absorption coefficient.

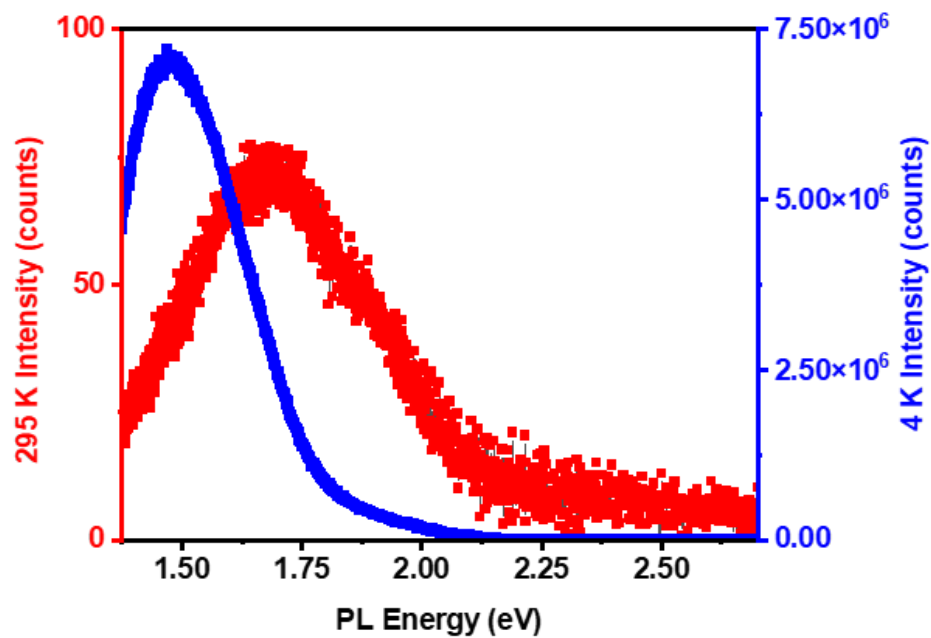

**Supplementary Figure 3: Room Temperature and Low Temperature PL.** A comparison between the room temperature PL (red) and 4 K PL (blue). The PL intensity of the 4 K spectra is adjusted to match the room-temperature spectra's excitation power and collection time. Room temperature PL in SbPS$_4$ is nearly 5 orders of magnitude less intense. The central energy of the PL peak for SbPS$_4$ at room temperature is shifted towards higher energy by approximately 250 meV, compared to its PL at 4 K.

**Supplementary Note 2: Theory of SbPS₄ Electronic Structure**

Supplementary Figure 4 shows the electronic structure calculations of $SbPS_4$ using DFT calculations employing norm-conserving pseudopotentials and utilizing the widely accepted Perdew-Burke-Ernzerhof (PBE) exchange-correlation functional[2]. To ensure accurate results, a rigorous sampling of the Brillouin zone of a $99 \times 2 \times 1$ Monkhorst-Pack k-point grid mesh was used. The convergence of the calculations was achieved by employing a plane wave basis set with an energy cut-off of 100 Ry. Supplementary Figure 4a displays the band structure of $SbPS_4$, while Figure 3b shows the total density of states (DOS) and projected DOS for the Sb, P, and S atoms, providing comprehensive insight into the electronic properties. The projected DOS analysis reveals that the valence band corresponds to the orbitals of the S atoms, while the conduction band includes contributions from all atomic orbitals. Supplementary Figures 4c-e depict DOS contributions from Sb, P, and S atoms, respectively, and their decomposed contributions from individual orbitals. It displays the contribution of each atomic orbital to the DOS of the structure.

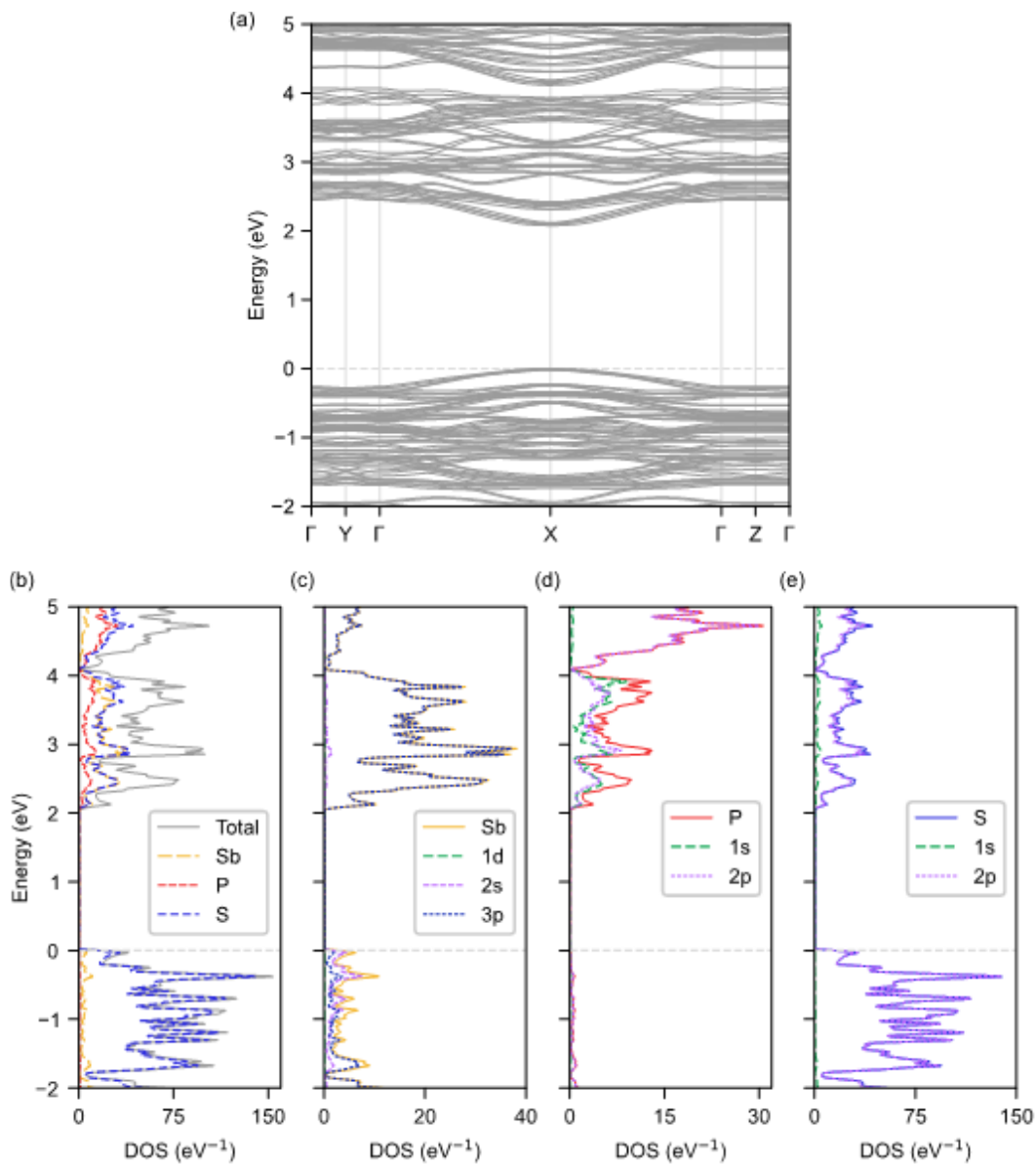

**Supplementary Figure 4: Electronic structure calculations of SbPS₄.** (a) Band structure of SbPS4. (b) Total density of states (DOS) and projected DOS for Sb, P, and S atoms. (c-e) DOS contributions from Sb, P, and S atoms, respectively, and their decomposed contributions from individual orbitals (1s, 1d, 2p, and 3p). The DFT calculations used the norm-conserving pseudopotentials for Pb and Sb atoms and the Perdew-Burke-Ernzerhof (PBE) exchange-correlation functional. The plane wave energy cut-off was set to 100 Ry, and the Brillouin zone integration was performed using a 99x2x1 Monkhorst-Pack k-point grid. The maximum energy of valence band is set to zero.

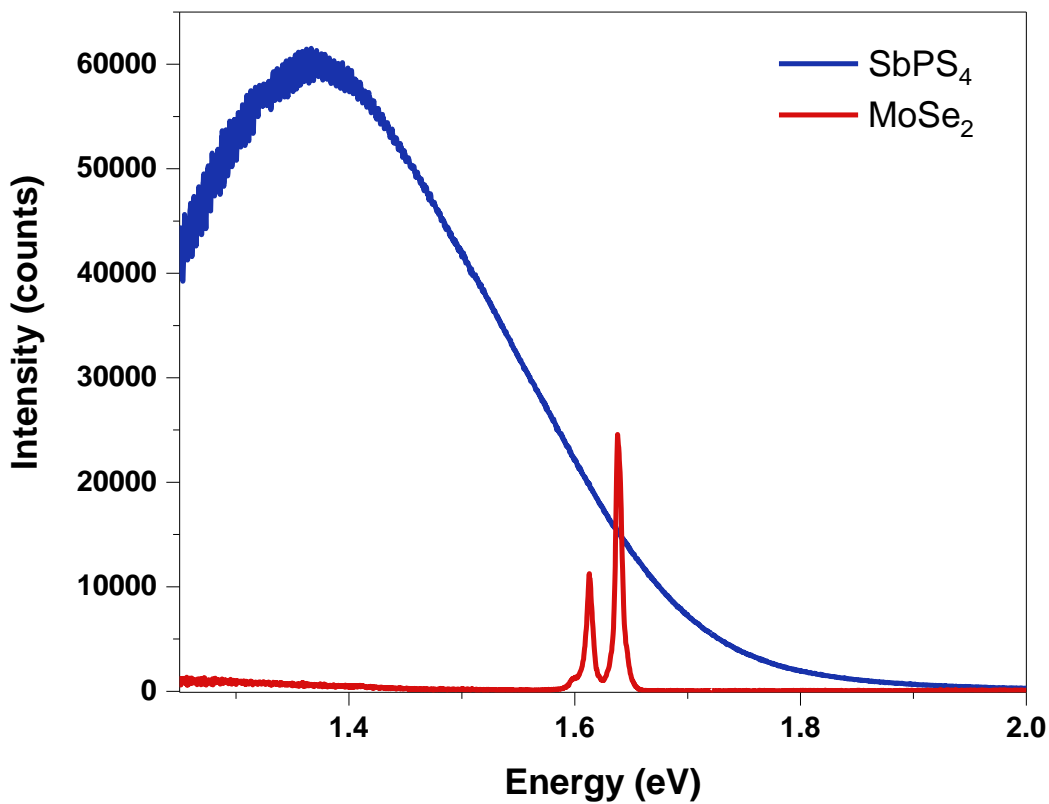

**Supplementary Figure 5: PL Comparison of SbPS₄ and hBN-encapsulated MoSe₂.** PL of a SbPS₄ nanobundle and hBN-encapsulated MoSe₂ monolayer while both materials were subjected to identical measurement parameters: a 405 nm laser with 1 μW power and a 5-second exposure time in the same optical setup at 5 K. In order to compare their intensity, we divided the spectrally integrated counts by the sample thickness, which here was ~200 nm for the SbPS₄ and 0.7 nm for monolayer MoSe₂. The SbPS₄ is measured at $4.5 \times 10^6$ Counts/nm, whereas MoSe₂ emits at a rate of $2.8 \times 10^7$ Counts/nm.

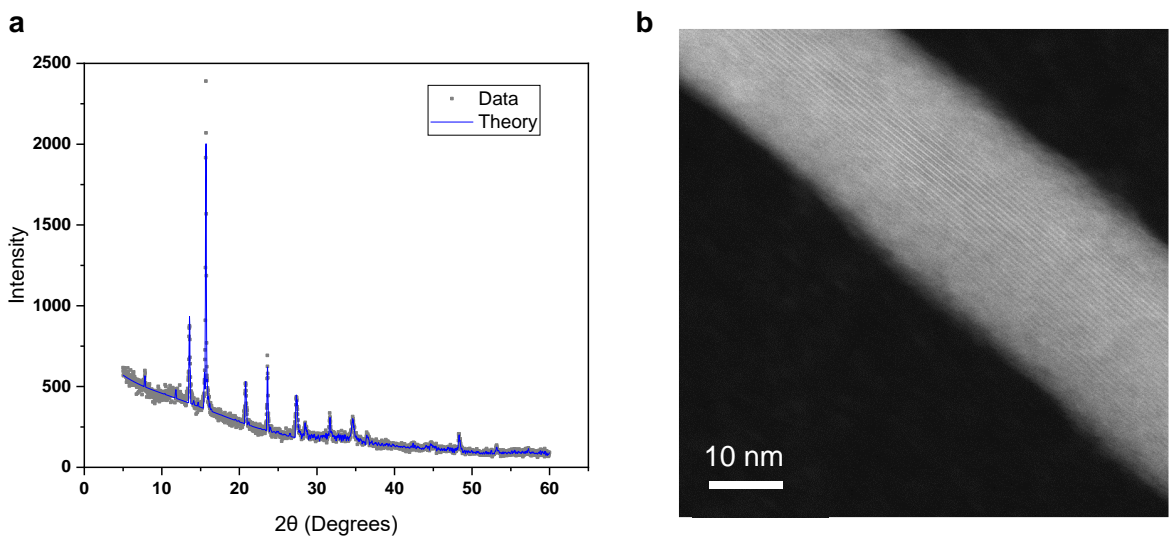

**Supplementary Figure 6: X-ray diffraction and transmission electron miscopy of SbPS₄** (a) Powder x-ray diffraction of SbPS₄. The grey points show the measured intensity and the blue curve is the calculated intensity using Le Bail method. The overall diffraction data is consistent with the reported crystal structure[3,4]. (b) Transmission electron microscope (TEM) image of a SbPS₄ nanobundle. The observation of van der Waals bonded tubes agrees well with previous reports[3].